\documentclass[twocolumn,twoside,10pt,nodate]{article}
\usepackage{amsmath,amssymb}
\usepackage{fancyhdr}
\usepackage{graphicx}% Include figure files
\usepackage[english]{babel}
\usepackage{hyperref}

\newcommand{\titolo}[2]{\title{\Large\bf\vspace{-1cm}#1}}

\begin{document}

% ** Please do NOT remove/change anything above this line ****

\titolo{2013-2016 review: HE Neutrino and UHECR Astronomy?}

% ** Please do NOT remove/change anything above this line ****
\maketitle
\thispagestyle{fancy} % ** do NOT remove this line **
 Since recent $2013$ the energy neutrino detector, ICECUBE, detected a flavor change in highest tens TeVs up to PeVs $\nu$ events. The atmospheric muon neutrino tracks in ICECUBE
 above $60$ TeV are suddenly overcome by (probable electron or tau neutrino) shower and cascades. Such an energy neutrino signal change was generally indebted to a new astrophysical HE neutrino component [1]. But, no known $\gamma- X$ source candidate has been associated to the observed $\nu$ ICECUBE $54$ events. A prompt atmospheric neutrino might be also polluting making the flavor change: the absence of any tau (a double bang imprint) presence in a dozen of highest $200$ TeV events, might favor, such a noisy prompt neutrino role. Indeed, unfortunately: No tau imply a non guaranteed neutrino astronomy [1]. Anyway most (4:1)  of those $\nu$ ICECUBE $54$ events are cascades (spherical lightening in ICECUBE) with a poor ($\pm 10^{o}$) directionality. Rarer, a dozen, muon HE tracks are more precise but too few. No obvious astronomical $\gamma\leftrightarrow \nu$ correlation have been found even by contained muon tracks. We proposed as astronomical neutrinos pointers the highest $\nu_{\mu} \mapsto \mu$ tracks made by crossing muon, born outside the ICECUBE detector [2]. Indeed such a HE $\mu$ tracks originated by  primary muon neutrino interactions outside the cubic km volume may cross upward in a larger volume, offering a more abundant rate and a sharper map. We foresaw that such crossing muons as the leading tool to a soon HE neutrino astronomy; self clustering events, in particular in connection with rarest narrow UHECR (Ultra High Energy Cosmic Rays) clustering may discover the sources [2]. Very recent $\mu$ tracks data  might be confirming our earliest hinted sources[2].
Indeed the UHECR  signals may be a complementary astrophysical imprint of HE $\nu$; they are revealed in ground largest array detectors, as AUGER and TA (Telescope Array) or AGASA, discovering a large numbers (hundreds) of such UHECR events. Because of their energy UHECR, if nucleons,  may fly almost unbent: they may trace sharply their sources within a bounded (one percent) GZK of Universe radius. This GZK limit exist because of the cosmic radiation opacity to UHECR $p$ by  photo-pion opacity. Lightest nuclei are even more constrained and they bend and smear at ($\pm 10^{o}-15^{o}$) angles. Heavier UHECR are mixed and lost by largest bendings. In UHECR maps we recognized three main results: 1, the remarkable absence of (otherwise expected for $p$ UHECR) of Virgo source (the nearest and largest cluster of galaxy in GZK universe); 2, a twin Hot Spot of UHECR clustering (North and South ones) whose smeared origin might be  associated (by us) to lightest nuclei bending, (around Cen A and M82), the nearest AGNs; 3, the consequent lightest nuclei nature in UHECR, that could stop at shortest distances (a few Mpc photo-nucleon dissociation) explaining the mystery of Virgo (20 Mpc) absence. These compositions have been confirmed by later and last (2015-2016) AUGER shower simulations, favoring light and lightest nuclei UHECR [3]. Finally among the amazing HE $\nu$ puzzles  in ICECUBE data is the (unexpected) absence of any correlation between the HE (TeVs) muon tracks and the GRBs ($\gamma- X$) events. This absence make the best celebrated GRB HE $\nu$ candidate, based on  GRB Fireball hadronic models, no longer a HE $\nu$  source. To overcome the neutrinos absence in GRB, we considered a new version of  GRB made by a precessing gamma blazing jet, fed mostly by a pure primary electron pairs jet powering a $\gamma$ jet by Inverse Compton Scattering. The model is born within an empty NS-NS (NS, Neutron Star) or NS-BH, (BLack Hole), binary system merging at late spiral and collapse times. We foresee that the lightest companion (NS) suffer by tidal stripe tease (by BH) of mass and its neutron escape and decay, feeding an accretion disk at BH, make a disk mainly dominated by proton and a charged one. The relativistic electron via neutron decay are forced by the axial magnetic field to shrinkage along their axis, accelerating and feeding a GeVs-TeVs  $e^{-}$ as well as a $e^{\pm}$ jet and a consequent collinear $\gamma$ jet: its precessing blazing are seen by far observer as GRBs.  The late survived NS  fragment (before a final merging on BH) may sometimes lead (below $0.2$ solar mass) to the NS instability and its sudden explosion, that mimic a SN explosion. This explain  the rare (a few percent) and mysterious  presence of a consecutive (to GRB burst)  (apparent) SN-like event. The new model may also explain by an early off-axis  blazing, of the  $\gamma$ jet, sudden  appearance, of the (otherwise mysterious) X-gamma precursor to GRB whose weak flash before the main GRB is unreasonable in any Fireball models; NS-NS systems may also follow similar scenarios [4].
\begin{figure}[h]
\centering
\includegraphics[width=0.40\textwidth,angle=0]{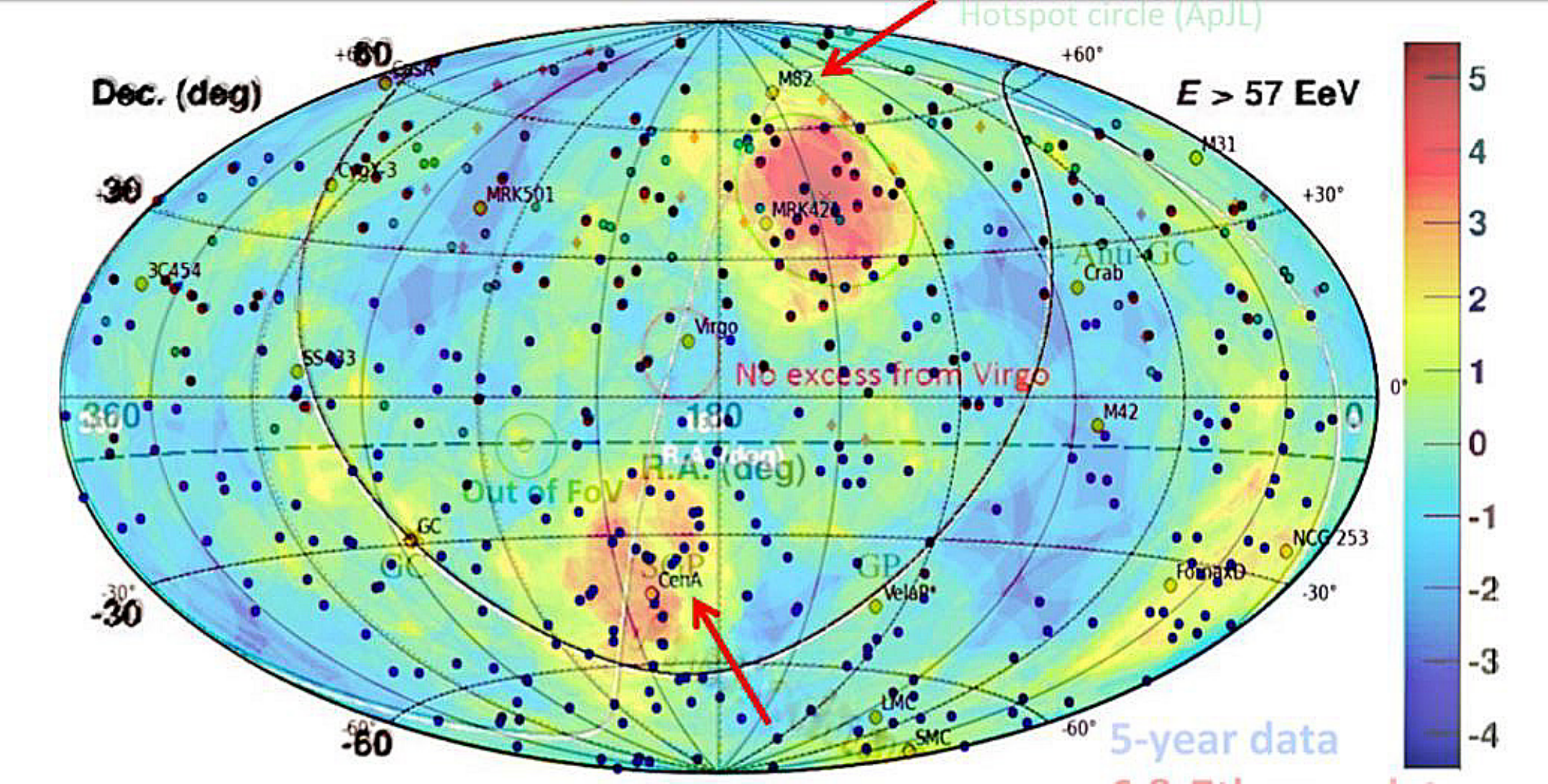}
\caption{\small
The two main smeared Hot Spot North,South, and the M82 and Cen A tagged sources.} % Please do NOT change the caption style
\end{figure}
%%%%%%%%
% Please do NOT remove the following two lines
% \\
% \\
%%%%%
\vfill
\small{
\noindent
\textbf{References}
\\
1. D.~Fargion, et al, NIMA \ \textbf{753}, 9-13, (2014); PoS FRAPWS2014 \ \textbf{028} ,(2016)\\
2. D.~Fargion, P.Oliva, Nuclear Physics B,(Proc.Suppl.)\ \textbf{213-217}, 256-257,(2014); \ \textbf{279-281},198--205,(2016);
   \ \ arXiv:1611.00079; (2017);\\
3. D.~Fargion,, et al, EPJ Web Conf.\ \textbf{99}  08002; (2015);\\
4.  D.~Fargion, P.Oliva, Nuclear Physics B, in press; \ \textbf{arXiv:1605.00177} (2017)\\
%%%%
% Please do NOT remove the following two lines
\textbf{Daniele Fargion; Physics Depart, Rome1 and INFN, Italy}
}
\end{document}